# On the structure and electronic properties of $Fe_2V_{0.8}W_{0.2}Al$ thin films

E. Alleno[1*], A. Berche[2], J.-C. Crivello[1], A. Diack Rasselio[1], P. Jund[2*]

1 Univ Paris Est Créteil, ICMPE, UMR CNRS UPEC 7182, 2 rue Henri Dunant, F94320 Thiais, France

2 ICGM, Univ. Montpellier, CNRS, ENSCM, Montpellier, France

Corresponding authors: alleno@icmpe.cnrs.fr, philippe.jund@univ-montpellier.fr

A very large thermoelectric figure of merit $ZT$ = 6 at 380 K has recently been reported in $Fe_2V_{0.8}W_{0.2}Al$ under thin-film form (Hinterleitner et al., Nature 576 (2019) 85). Under this form, $Fe_2V_{0.8}W_{0.2}Al$ experimentally crystallizes in a disordered $A$2 crystal structure, different from its bulk-form structure ($L2_1$). First principles calculations of the electronic structure performed in $A$2-$Fe_2V_{0.8}W_{0.2}Al$ supercells generated by the Special Quasi-random Structure (SQS) method are thus reported here. These calculations unambiguously indicate that $A$2-$Fe_2V_{0.8}W_{0.2}Al$ is a ferromagnetic metal at 0 K, displaying a small Seebeck coefficient at 400 K (< 30 µV K$^{-1}$). The present results contradict the scenario of the occurrence of a deep pseudo-gap at the Fermi level, previously invoked to justify $ZT$ = 6 in $Fe_2V_{0.8}W_{0.2}Al$ thin films.

Thermoelectric devices are heat-to-energy conversion systems, free of moving parts, position-independent, and operating in a wide range of temperatures. They are considered for applications such as recovering heat wasted in industrial activities or for powering the numerous wireless sensors required by an application such as the "Factory 4.0" [1]. Their conversion efficiency is an increasing function of the dimensionless figure of merit $ZT$ of their constituting materials. $ZT$ is defined by the relation $ZT = \frac{\alpha^2 T}{\rho \lambda}$, where $\alpha$ is the Seebeck coefficient, $T$ the absolute temperature, $\rho$ the electrical resistivity, and $\lambda$ the thermal conductivity.

The current state-of-the-art thermoelectric material at 300 K is $Bi_2Te_3$ [2], which displays a value of $ZT$ = 1 at this temperature. Unfortunately, it is constituted by expensive and toxic tellurium, these being some of the reasons preventing its spread to a larger market. $Fe_2VAl$ is a promising Heusler alloy since its elemental constituents are cheap [3] and non-toxic. When doped, it displays a thermoelectric power factor - $PF = \alpha^2/\rho \geq 5$ mW m$^{-1}$ K$^{-2}$ - larger than in $Bi_2Te_3$ [4, 5]. This feature could be related to a paramagnetic semi-metallic ground state (pseudo-gap ~ 0.2 eV), as obtained by Density Functionnal Theory (DFT) band structure calculations based on the Generalized Gradient Approximation (GGA) [6-9]. In addition, recent DFT-GGA calculations have shown that its electronic band gap increases with temperature, thus mitigating the otherwise unfavorable bipolar effect [10]. However, its thermoelectric performances are strongly limited by its too large thermal conductivity, roughly one order of magnitude larger than in $Bi_2Te_3$ [11], leading to $ZT$ ~ 0.1 at 300K.

Materials in low-dimensional form like thin films, quantum wells or nanowires are now well-known to display improved thermoelectric performances [12, 13]. In their recent article entitled "Thermoelectric performance of a metastable thin-film Heusler alloy", B. Hinterleitner et al. [14] have indeed reported an exceptionally large $ZT$ of 6 at 380K in a $Fe_2V_{0.8}W_{0.2}Al$ thin film. $ZT$ = 6 is nearly two orders of magnitude larger than in $Fe_2V_{0.9}W_{0.1}Al$ synthesized in bulk form by Mikami et al. ($ZT$ = 0.1 at 350K) [15]. This very large dimensionless figure of merit arises from the large Seebeck coefficient $\alpha$ = - 550 µV K$^{-1}$ reported by the authors, combined with a small value of the electrical resistivity ($\rho$ = 6 µΩ m), leading to an unexpectedly large power factor *PF* = 50 mW m$^{-1}$ K$^{-2}$.

Based on the Powder X-ray Diffraction (PXRD) data shown in ref. [14], it can easily be verified that the crystal structure of the Fe$_2$V$_{0.8}$W$_{0.2}$Al thin film is not the ordered *L*2$_1$ structure in which it crystallizes in its bulk form [16], but a disordered body-centered cubic structure (*bcc*, labelled *A*2 in strukturbericht), where all the atoms share the same crystallographic site. This structure is most probably metastable and stabilized by the strains caused by the underlying Si (100) substrate on which the film is sputtered. This point will be further commented after the presentation of the calculated formation energies.

This change in the crystal structure of the thin film leads to changes in its electronic structure, which were calculated by Hinterleitner et al. in the framework of DFT-GGA. The calculated density of states (DOS) and band structure are indeed key results in ref. [14] to discuss the origin of the large reported power factor. According to the authors, the large Seebeck coefficient could be ascribed to the very large slope in the DOS at the Fermi level whereas the small resistivity value would be related to the occurrence close to the Fermi level of highly mobile "Weyl-like fermions" "protected from backscattering by non-magnetic disorder and defects".

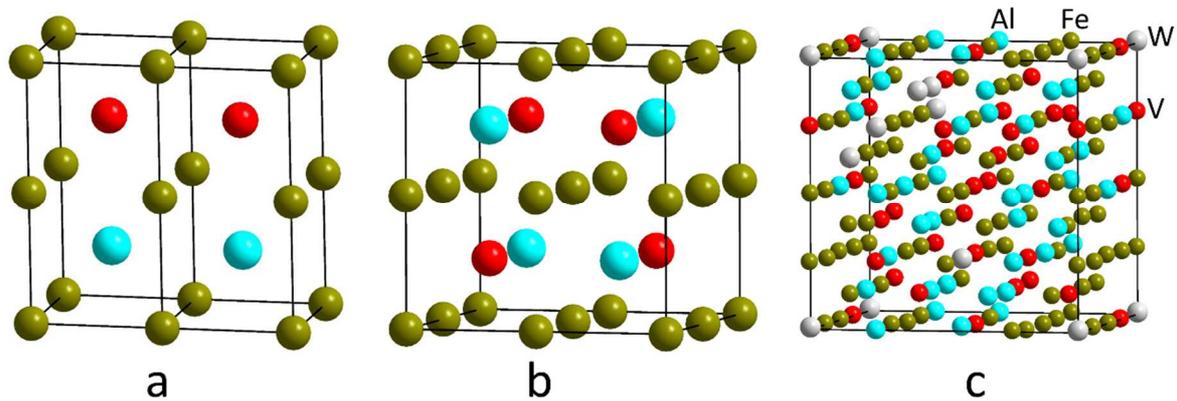

Fig. 1. a. Tetragonal unit-cells derived from the figure "Extended data Fig. 9" in ref. [14]. b. *L*2$_1$ structure represented by its cubic unit-cell (space group *Fm*-3*m*). c. One of the two *A*2 supercells generated by the SQS procedure (see Appendix for details): 2 × 2 × 2 supercell containing 128 atoms. Blue spheres: Al atoms, khaki spheres: Fe atoms, red spheres: V atoms, light grey spheres: W atoms.

However, to calculate the electronic structure of the Fe$_2$V$_{0.8}$W$_{0.2}$Al thin film, Hinterleitner et al. made use of a crystal structure that is not correct. In Fig. 1, it is obvious that their "tetragonal" unit cell (Fig. 1a) describing the Fe$_2$VAl parent compound differs from the cubic *L*2$_1$ structure (Fig. 1b): in the "tetragonal" unit cell the V and Al atoms do not share the same (100) planes while in the actual *L*2$_1$ structure, they do [17]. The 80 atoms supercell build from the "tetragonal" sub-cell to describe the disordered body-centered structure of the thin film and shown as inset in Fig. 2a, is thus wrong from a crystallographic point of view. It is also structurally wrong since it is not disordered, conversely to the authors' experimental result. To better show this, we calculated the theoretical PXRD pattern of the 80 atoms "tetragonal" supercell (Fig. 2a) and compared it to the experimental PXRD pattern reported in ref. [14] and displayed in Fig. 2b. The most salient differences consist in low angle lines (2$\theta$ < 40 °) that are absent in the experimental pattern (Fig. 2b) while occurring in the calculated one (Fig. 2a). According to the literature on the structures of Heusler alloys [18] and Fe$_2$VAl [17], these extra lines can easily be ascribed to the absence of disorder in the 80 atoms supercell of ref. [14]. It can hence be surmised that the calculated band structure reported in ref. [14] does not represent the one of the synthesized thin film. To verify this, we carried out DFT calculations of the electronic structure in disordered *bcc-A2* Fe$_2$V$_{0.8}$W$_{0.2}$Al, which is the structure of the thin film of ref. [14].

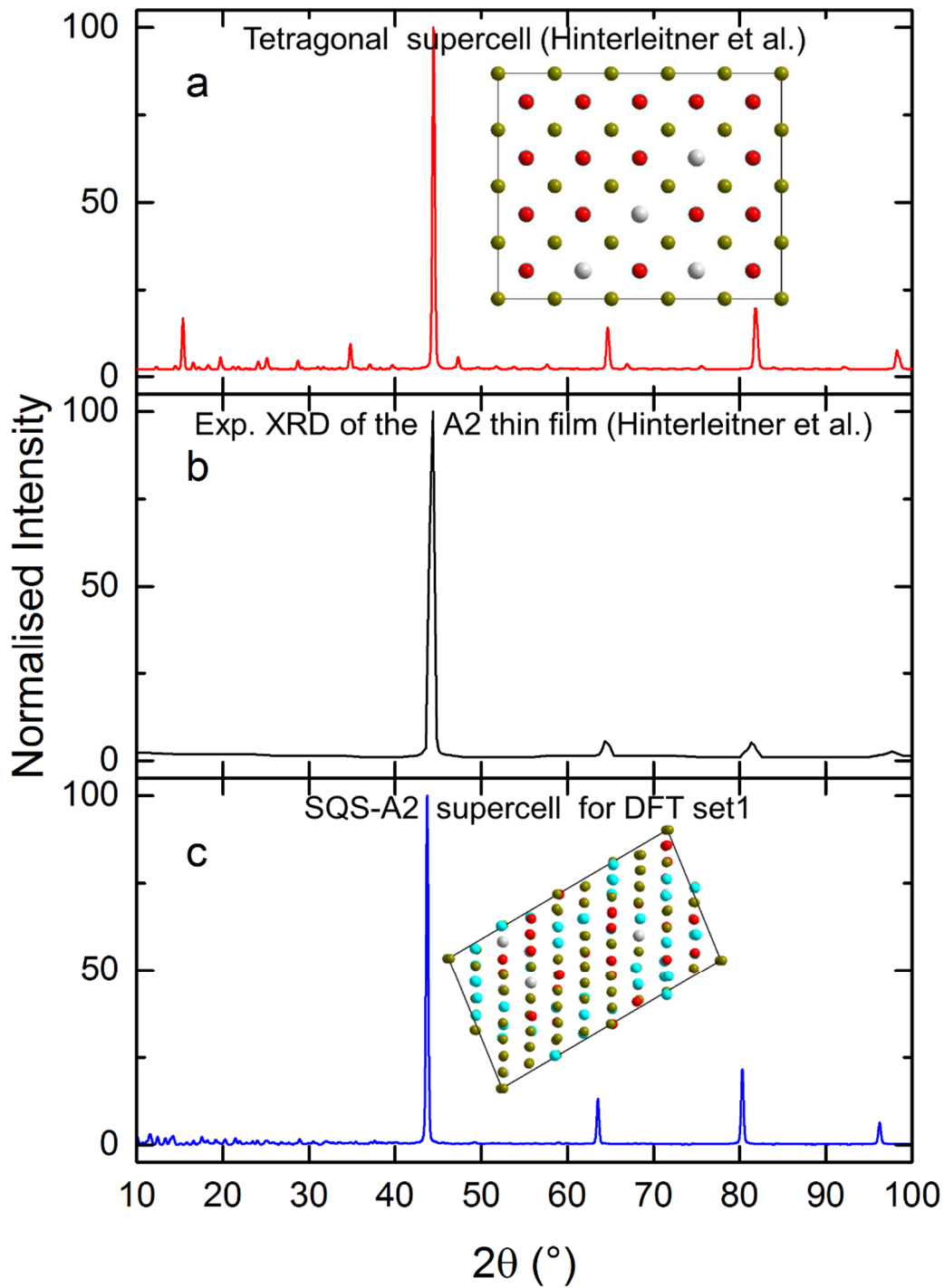

Fig. 2. a. Calculated PXRD pattern of the 80 atoms "tetragonal" supercell used in ref. [14] for DFT calculations of their Fe$_2$V$_{0.8}$W$_{0.2}$Al thin film (shown in the inset); b. Experimental PXRD pattern of the Fe$_2$V$_{0.8}$W$_{0.2}$Al thin film from ref. [14]; c. Calculated PXRD pattern of the 160 atoms SQS-*A2* supercell for DFT calculations of the Fe$_2$V$_{0.8}$W$_{0.2}$Al thin film (set1, shown in the inset). Every atom displayed with the same color code as in Fig. 1. All the PXRD patterns are normalized by setting the most intense line to 100.

The first necessary step is to build a supercell which is representative of the considered disorder. For this purpose, the Special Quasi-random Structure (SQS) method [19] helps to choose a reasonable cell size representative of atoms randomly distributed on a chosen specific crystallographic site. Considering the $Fe_2V_{0.8}W_{0.2}Al$ composition, several options of disorder are available. The first intuitive one is based on the $L2_1$ structure ($Fm$-$3m$) of the $Fe_2VAl$ host prototype, where the mixing of 0.8 V + 0.2 W is obtained on the single $4a$ site (Fig. 5 in the Appendix). The second option is the mixing of all atoms on all sites, leading to the *bcc* or *A*2 structure. To be as rigorous as possible, the electronic structure of the disordered $L2_1$ or *bcc-A*2 crystal structures has been calculated independently by the two distinct research groups involved in this study (set 1 and set 2). In both cases, the choice of the atomic arrangement for the two proposed structures ($L2_1$ and *A*2) has been performed using the SQS methodology as described in detail in the Appendix. Two sets of data have thus been obtained on several supercells (for instance the *bcc-A*2 $Fe_2V_{0.8}W_{0.2}Al$ set1 and set2 displayed in Fig. 2c and Fig. 1c respectively), with various SQS and DFT parameters, which allow to discard a methodological bias on the present results. DFT calculations have been performed within the GGA, similarly to what has been done by Hinterleitner et al. More details about the DFT calculations are also provided in the Appendix.

The calculated PXRD pattern of one of the generated SQS *bcc-A*2 $Fe_2V_{0.8}W_{0.2}Al$ (set1) is displayed in Fig. 2c and it can easily be seen that it looks like the experimental one (Fig. 2b) since its low angle extra lines ($2\theta < 40$ °) are much less intense than in the calculated PXRD pattern (Fig. 2a) of the "tetragonal" supercell used in ref. [14]. These SQS supercells provide thus a structural representation much closer to the reality of the experimental $Fe_2V_{0.8}W_{0.2}Al$ thin film than the ordered "tetragonal" supercell.

The formation energies of pure and W-doped $Fe_2VAl$ are provided in Table 1. The DFT results explicitly show that whatever the W content, the $L2_1$ structure is the most stable phase. The "tetragonal" structure used by Hinterleitner et al. is 0.05 eV.atom$^{-1}$ less stable than the $L2_1$ structure while *bcc-A*2 is clearly metastable at 0 K (difference in energy with $L2_1$ around 0.25 eV.atom$^{-1}$). This energy difference questions why the metastable *bcc-A*2 structure is observed in the thin film. Again, only strains induced by the silicon substrate or specific epitaxial growth conditions can explain the occurrence of the *bcc-A2* structure in the experimental PXRD pattern (Fig. 2b).

| $\Delta_f E$ (eV.at$^{-1}$) | Pristine $Fe_2VAl$ | | $Fe_2V_{0.8}W_{0.2}Al$ | |
|---|---|---|---|---|
| | Set 1 | Set 2 | Set 1 | Set 2 |
| $L2_1$ | -0.426 | -0.432 | -0.409 | -0.381 |
| "tetragonal" | -0.384 | -0.387 | -0.339 | -0.340 |
| *bcc-A*2 | -0.068 | -0.149 | -0.139 | -0.136 |

Table 1: Formation energy of pristine and W-doped $Fe_2VAl$ for various crystal structures.

The densities of states of the "tetragonal" and SQS *bcc-A*2 structures, all calculated in the present work, are shown altogether in Figure 3, for an easy comparison with the work of Hinterleitner et al [14]. Similar to the DOS of $Fe_2VAl$ [20] and to the DOS of SQS-$L2_1$ $Fe_2V_{0.8}W_{0.2}Al$ (Appendix, Fig. 6), the electronic structure of "tetragonal" $Fe_2V_{0.8}W_{0.2}Al$ presents a pseudo-gap at the Fermi level (Fig 3a). The additional electrons coming from the W substitution leads to a complete filling of the valence band and to a partial filling of the conduction band. In agreement with the result obtained in ref. [14], "tetragonal" $Fe_2V_{0.8}W_{0.2}Al$ is a *n*-type doped semi-metal. Conversely, the DOS of the *bcc- A*2 structure (set 1) displayed in Fig. 3b does not present a pseudo-gap but rather exhibits an unambiguous metallic character, similarly to the host *A*2-$Fe_2VAl$ [17] or to partially disordered $Fe_2VAl$ [8]. Moreover, spin-dependent calculations strongly suggest it is a ferromagnet at 0 K ($m$ = 0.76 µB / at).

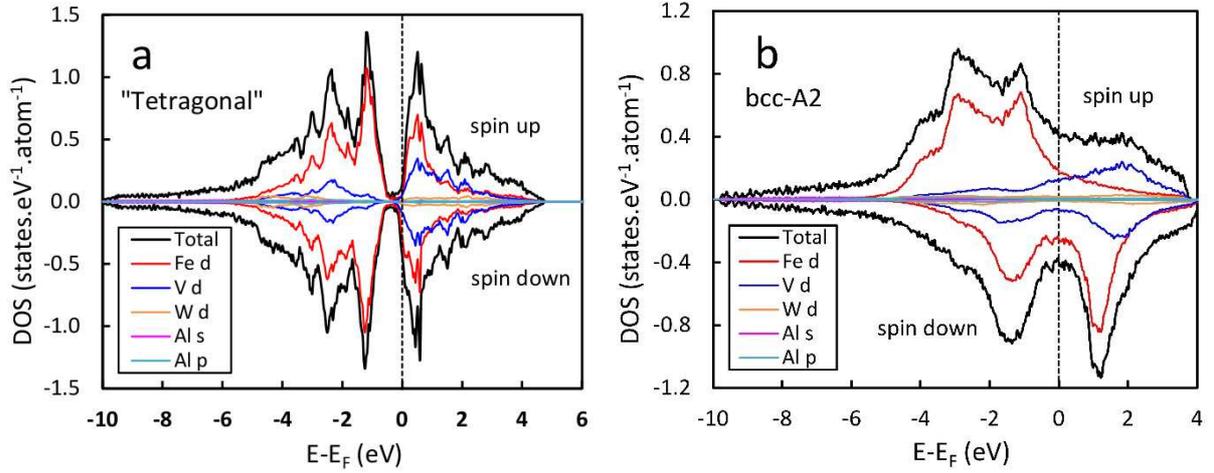

Fig 3. a. DOS of $Fe_2V_{0.8}W_{0.2}Al$ in the "tetragonal" structure, re-calculated from the crystallographic structure presented in ref. 14, similarly to the original representation; partial DOS corresponding to the Fe-*d*, V-*d*, W-*d*, Al-*s* and Al-*p* orbitals have been included. b. DOS for the *bcc-A*2 structures generated by the SQS method (set 2); partial DOS corresponding to the Fe-*d*, V-*d*, W-*d*, Al-*s* and Al-*p* orbitals have been included. By convention, spin up DOS are positive while the spin down DOS are negative.

Such a metallic DOS is not compatible with "the occurrence close to the Fermi level of the highly mobile Weyl-like fermions protected from backscattering by non-magnetic disorder and defects", invoked by Hinterleitner et al. to justify the small resistivity observed in $Fe_2V_{0.8}W_{0.2}Al$ thin films. It is neither compatible with the Seebeck coefficient of $-550 \mu V.K^{-1}$ experimentally measured at 380 K in ref. [14] and ascribed to the very large slope in the DOS at the Fermi level. To better convince the reader, we made use of the BoltzTraP software [21] to calculate from the electronic structure the Seebeck coefficient (Fig. 4) in *bcc-A*2 $Fe_2V_{0.8}W_{0.2}Al$ at 400K. It is obvious that whatever the value of the chemical potential around the calculated Fermi level, a maximum absolute value of $\pm 30 \mu V.K^{-1}$ is predicted in agreement with the metallic ground state of $Fe_2V_{0.8}W_{0.2}Al$.

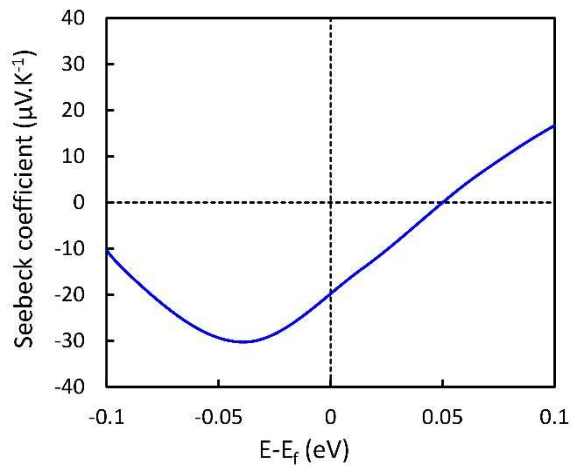

Fig. 4: Seebeck coefficient calculated as function of the chemical potential referred to the Fermi level for the $Fe_2V_{0.8}W_{0.2}Al$ compound in the *bcc-A*2 structure at 400K (set 2)

To summarize, Hinterleitner et al. performed DFT calculations on an ordered "tetragonal" supercell with the goal of providing firm theoretical grounds to their experimental observation of $ZT = 6$ in a $Fe_2V_{0.8}W_{0.2}Al$ thin film. However, we have shown that the "tetragonal" structure used for their calculations does not represent the experimentally established *bcc-A*2 structure of their disordered film. We thus calculated the electronic structure of disordered $Fe_2V_{0.8}W_{0.2}Al$ using SQS-generated supercells. These calculations indicate that this material is a ferromagnetic metal at 0 K, displaying a Seebeck coefficient that cannot exceed ± 30 µV K$^{-1}$ at 400K. The theoretical scenario invoked by Hinterleitner et al. to explain their $ZT = 6$ should hence be fully reconsidered. Finally, the present work also raises questions about their experimental results and could improve the awareness of experimentalists envisaging new research on the thermoelectric properties of thin film $Fe_2V_{0.8}W_{0.2}Al$.

**ACKNOWLEDGEMENTS**

The authors acknowledge the funding of this work by the "Agence Nationale pour la Recherche" through the contract "LoCoThermH" (ANR-18-CE05-0013-01). Set 1 calculations were performed using HPC resources from GENCI–CINES (Grant 2019–096175)

**APPENDIX: details on DFT calculations**

The choice of supercells corresponding to the two proposed mixtures ($L2_1$ and $A2$, see Fig. 5) has been managed using the *mcsqs* tool from the *ATAT* package [22]. The approach consists in generating supercells of $N$ atoms presenting neighbor correlation functions over the $q$ cluster, $\zeta_q$, which are equal or as close as possible to the ones of a fully random structure, *i.e.* choosing a configuration where the average correlation functions over a set of i$q$ clusters as : $<\zeta_q^{SQS}> \cong <\zeta_q^{random}>$. The error on the generation is estimated by the root-mean-square (RMS): $\sqrt{\sum_q \left(\zeta_q^{SQS} - \zeta_q^{random}\right)^2}$. According to the experience of SQS cells generated in a *bcc* phase [23], at least 4 first neighbors' pairs ($q = 2$) need to be considered.

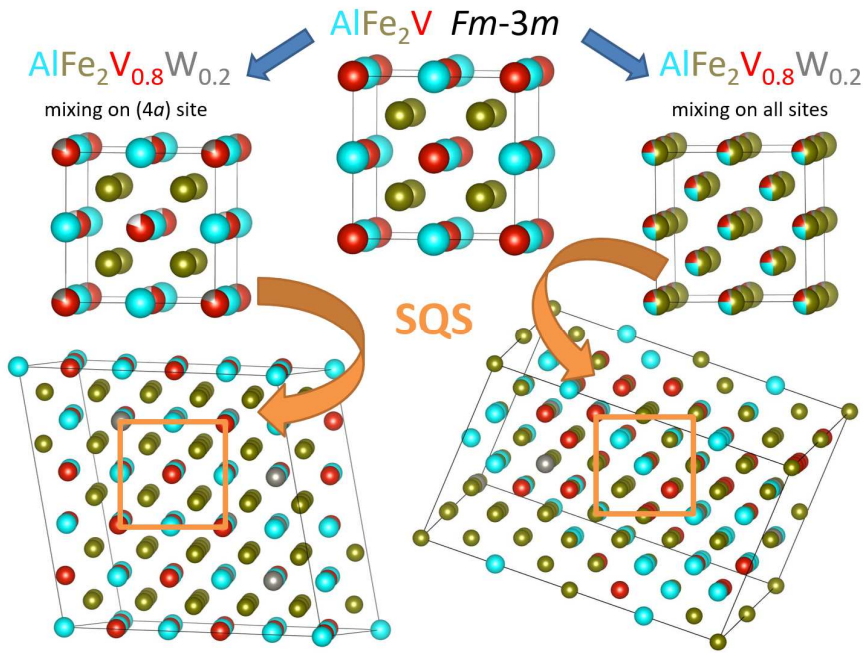

Fig. 5. Atomic distributions of disordered $Fe_2V_{0.8}W_{0.2}Al$ generated by SQS (set 1).

In the first set of calculations, the following clusters have been selected after several tests: 6 pairs ($q = 2$) + 3 triangles ($q = 3$) + 1 tetrahedron ($q = 4$) and 6 pairs + 1 triangles + 1 tetrahedron for $Fe_2V_{0.8}W_{0.2}Al$ in the $L2_1$ and $A2$ structures respectively. This leads to include the pair interactions up to 10 and 6 Å respectively, with a global RMS error below 0.08. This exhaustive search finds that SQS cells with $N = 160$ satisfy this criterion (with 8 W atoms / cell with a global composition $Fe_2V_{0.8}W_{0.2}Al$) for both the $L2_1$ and $A2$ structures. The lattice vectors and atomic positions of the SQS cells are respectively:

$$\begin{pmatrix} 0 & 3/2 & 1/2 \\ 0 & -1/2 & 5/2 \\ 5/2 & 0 & 1/2 \end{pmatrix} \begin{pmatrix} -3 & -1 & 1 \\ -1/2 & 3/2 & -7/2 \\ 2 & -6 & -2 \end{pmatrix}$$

The electronic structure has been subsequently calculated in the framework of the DFT using the all-electron projector augmented wave method [24] within the GGA of Perdew et al. [25] (PBE), as implemented in the Vienna Ab Initio Simulation Package (VASP) [26]. A plane wave basis set with 400 eV as cutoff energy has been chosen using spin polarization. The irreducible Brillouin zone (BZ) was sampled using $10 \times 6 \times 6$ and $8 \times 7 \times 4$ $k$-point meshes.

In the second set of calculations, supercells of the primitive cell of the $L2_1$ structure ($3 \times 3 \times 3$ containing 108 atoms including 6 W with a global composition $Fe_2V_{0.7778}W_{0.2222}Al$) and of the conventional $A2$ structure ($4 \times 4 \times 4$ containing 128 atoms including 6W with a global composition $Fe_2V_{0.8125}W_{0.1875}Al$) have been considered (see Fig. 1c). For the SQS, more triangles and tetrahedra have been taken into consideration and the following clusters have been selected: 5 pairs ($q = 2$) + 4 triangles ($q = 3$) + 4 tetrahedrons ($q = 4$) and 6 pairs + 4 triangles + 4 tetrahedrons for the $Fe_2V_{0.8}W_{0.2}Al$ in the $L2_1$ and $A2$ phases respectively. This leads to include the interactions up to 8 and 6 Å respectively. In these conditions, the global RMS errors are around 0.4, but the RMS error on the pairs are below 0.1, in agreement with the first set of calculations. DFT calculations were performed also using the code VASP within the PBE-GGA. The same standard versions of the PAW potentials for Fe ($3p^63d^64s^2$), V ($3s^23p^63d^34s^2$), W ($5d^46s^2$) and Al ($3s^23p^1$) as for set 1 were implemented. The first BZ was sampled using $3 \times 3 \times 3$ Monkhorst-Pack $k$-point meshes. The cutoff energy was set to 500 eV for these calculations. Both cell parameters and atomic positions were relaxed before electronic calculations within an energy accuracy of 1 μeV and 10 μeV/Å for the forces.

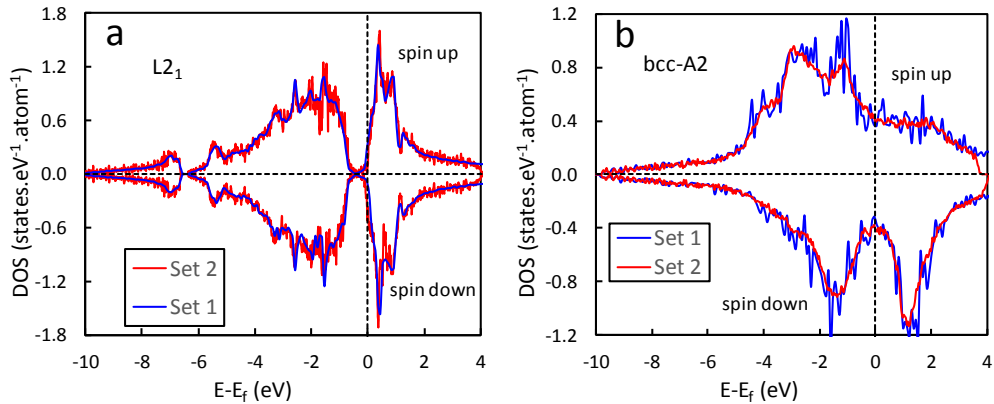

Fig. 6. Density of states of $Fe_2V_{0.8}W_{0.2}Al$ in the: a) $L2_1$ structure (V and W mixing on the $4a$ site) generated by SQS; b) bcc-$A2$ structure (Fe, V, W and Al mixing on the $2a$ site generated by SQS)